\documentclass[aps,prl,twocolumn,showpacs,superscriptaddress]{revtex4-2}

\usepackage{graphicx}  
\usepackage{amsmath}   
\usepackage{amssymb}   
\usepackage{color}
\usepackage[T1]{fontenc} 
\usepackage[dvipsnames]{xcolor}
\usepackage{hyperref}  
\hypersetup{
    colorlinks=true,
    linkcolor=blue,
    filecolor=blue,
    urlcolor=blue,
    citecolor=blue,
    pdfborder={0 0 0}
}

\newcommand{\apjs}{ApJS}

\newcommand{\apjl}{ApJL}
\newcommand{\aap}{A$\&$A}
\newcommand{\mnras}{MNRAS}

\begin{document}
\title{ A Star Cluster Population of High Mass   Black Hole Mergers in
  Gravitational Wave Data}

\author{Fabio Antonini}
\affiliation{Gravity Exploration Institute, School of Physics and Astronomy, Cardiff University, Cardiff, CF24 3AA, UK}
\author{Isobel M. Romero-Shaw} 
\affiliation{DAMTP, Centre for Mathematical Sciences, University of Cambridge, Wilberforce Road, Cambridge, CB3 0WA, UK}
\affiliation{Kavli Institute for Cosmology, Madingley Road, Cambridge, CB3 0HA, United Kingdom}
 \author{Thomas Callister}
\affiliation{Kavli Institute for Cosmological Physics, The University of Chicago, Chicago, IL 60637, USA}

\date{\today}

\begin{abstract}
Stellar evolution theories predict a gap in the black hole birth mass spectrum
as the result of 
pair instability processes in the cores of massive stars.
This gap, however, is not seen in  the binary black hole masses inferred from gravitational wave data. One explanation is that black holes form dynamically in dense star clusters where smaller black holes merge to form more massive black holes, populating the  mass gap.
We show that this model predicts a distribution of the effective and precessing spin parameters, $\chi_{\rm eff}$ and $\chi_{\rm p}$,  
 within the mass gap that is insensitive to assumptions about black hole natal spins {  and other astrophysical parameters}.
We analyze the distribution of 
$\chi_{\rm eff}$ as a function of primary mass for the black hole binaries in the third gravitational wave transient catalog. 
{  We infer the presence of
a high-mass and isotropically spinning population of black holes that is consistent with hierarchical formation in dense star clusters
and a pair-instability mass gap
with a lower edge at $44^{+6}_{-4} M_\odot$. We compute a Bayes factor $\mathcal{B}>10^4$ relative to models that do not allow for a high-mass population with a distinct 
$\chi_{\rm eff}$ distribution.}
Upcoming data will enable us to tightly constrain the hierarchical formation hypothesis and refine our understanding of binary black hole formation.
\end{abstract}


\maketitle

\noindent
{\it Introduction.}
Observations of gravitational waves (GWs) from binary black hole (BH) mergers provides an unprecedented window into the astrophysics of massive stars \cite{2015CQGra..32g4001L,2015CQGra..32b4001A,2016PhRvL.116f1102A,2019ApJ...882L..24A,2019PhRvX...9c1040A,LVKCollab2023,2021arXiv211103606T}. However, our ability to learn from these detections is hindered by uncertainties in the theory of 
massive binary star evolution and BH formation, and the dependence of model results on initial conditions and parameters \cite[e.g.,][]{2020PhRvD.102l3016A,2021AA...651A.100O,2021ApJ...910..152Z, 2022MNRAS.516.5737B}.
One of these uncertainties is the location (and presence) of a mass gap in the BH birth mass distribution due to {(pulsational) pair instability supernovae ((P)PISN)}, estimated to begin in the $\sim 40$--$70M_\odot$ range and ending at $\sim 130M_\odot$~\citep{2016A&A...594A..97B,2020ApJ...902L..36F,2020ApJ...897..100V,2021PhRvD.104d3015Z}.
Such a gap is not  seen in the BH mass distribution \cite{2021ApJ...913L..23E,LVKCollab2023}.
If such a gap exists, then the formation 
of BHs within the mass gap can  be explained
by scenarios involving dynamical interactions in dense star clusters or AGN disks, where BHs
 can grow through hierarchical mergers  \cite{2006ApJ...637..937O,2016ApJ...831..187A,2019PhRvD.100d3027R,2021ApJ...908..194T,2021MNRAS.505..339M,2023MNRAS.522..466A,2024arXiv240114837T,2024A&A...685A..51V}. 
The detection of this high mass  population will shed light on the origin of binary BH mergers as well as on the location of the (P)PISN mass gap.

The most precisely measured spin parameter from GW data is the effective inspiral spin $\chi_\mathrm{eff}$, a combination of the two component spins projected parallel to the orbital angular momentum~\cite{2008PhRvD..78d4021R}.
Previous work has shown that the distribution
of $\chi_{\rm eff}$ 
can give important insights
on the origin of the detected BH binary population
\cite{farr_distinguishing_2017,farr_using_2018,2021ApJ...922L...5C,2021MNRAS.507.3362T,2021NatAs...5..749G,2021ApJ...915L..35K,Wang:2022gnx,2022PhRvD.105l3024F,2022ApJ...935L..26F,2022ApJ...937L..13C,2023PhRvD.108j3009G,2024arXiv240403166R}.
The distribution of individual BH spins and
of the precessing spin parameter $\chi_{\rm p}$~\cite{PhysRevD.91.024043} have also been leveraged in attempts to identify a dynamically-formed population in the data
\cite{2017CQGra..34cLT01V,2022ApJ...928...75H,Wang:2022gnx,2020ApJ...894..129S,2022arXiv220303651M,2023arXiv230302973L,2024arXiv240409668L,2024arXiv240601679P}, and correlations between BH spins and binary mass ratio \cite{2021ApJ...922L...5C,2022MNRAS.517.3928A,2022ApJ...928..155T,2023ApJ...958...13A,LVKCollab2023,2024PhRvD.109j3006H} and possibly redshift \cite{2022ApJ...932L..19B,2024PhRvD.109j3006H} have been inferred from the data.
Other studies have examined possible relationships between BH mass and spin, as would be expected from a population of hierarchical mergers~\cite{2022ApJ...932L..19B,2022arXiv220303651M,2022ApJ...935L..26F,2022ApJ...928..155T,2023arXiv230401288G,2023arXiv230302973L,2024arXiv240403166R}; thus far, evidence for any correlation between BH masses and spins has been tentative, with different studies yielding different conclusions. A clear detection of the (P)PISN mass gap and a population of hierarchical mergers remains elusive.

Here, we consider the $\chi_\mathrm{eff}$ and $\chi_{\mathrm p}$ distributions as a function of primary BH mass $m_1$ to identify signatures of hierarchically formed BHs in the third gravitational wave transient catalog (GWTC-3).
\newline

\noindent
{\it Analytical model.} 
We consider the  spin parameters $\chi_{\rm eff}$ and $\chi_{\rm p}$ for a dynamically assembled BH merger population, in which the primary is a second-generation (2G) BH formed via an earlier merger and the secondary is a first-generation (1G) BH representing the direct end product of stellar evolution.
The effective inspiral spin is defined as 
$\chi_{\rm eff} = (m_1 a_1 \cos\theta_1 + m_2 a_2 \cos\theta_2)/(m_1 + m_2)$ \citep{Santamaria2010, ajith2011},
and the effective precessing spin parameter is 
$\chi_{\rm p} = \max \left(a_1 \sin\theta_1, \frac{3+4q}{4+3q} q a_2 \sin\theta_2 \right)$ \citep{hannam2014, schmidt2015},
where $q = m_2/m_1$ denotes the binary mass ratio, $a_1$ and $a_2$ are the dimensionless component spin magnitudes, and $\theta_1$ and $\theta_2$ are the angles between each spin vector and the orbital angular momentum.

For dynamical formation {in a dense star cluster}, we expect that
[i] spin and orbital angular momenta  are isotropically oriented \cite{Rodriguez2016c}. For 1G+2G mergers we also expect: 
    [ii]  $a_1=\tilde{a}\simeq 0.69$, in line with predictions from numerical-relativity simulations \cite{2008PhRvD..78d4002R};
    [iii] $a_1\gg a_2$, as a significant 1G+2G merger rate requires that BHs are formed with small spin, allowing for a higher cluster retention probability of merger remnants \cite{2019PhRvD.100d3027R,2023MNRAS.522..466A}; and
    [iv]  $m_1
    \simeq 2m_2$, based on dynamical selection favouring the high-mass end of the BH mass function for both BH primary progenitors and for the secondary BH \cite{1975MNRAS.173..729H}.

    Given the above considerations, we expect that the most likely values of the spin parameters for a  1G+2G mergers are
$\chi_{\rm eff}\simeq \frac{\tilde{a}}{{1.5}}\cos \theta_1$
and
$\chi_{\rm p}
\simeq \tilde{a}\sin\theta_1$,
leading to the  mean relation
$\chi_{\rm p}=\sqrt{\tilde{a}^2-\left(1.5\chi_{\rm eff}\right)^2}$. 
Isotropy  then implies 
a uniform distribution for $\chi_{\rm eff}$ within $|\chi_{\rm eff}| < {\tilde{a}}/{{1.5}}\simeq 0.47$, and  $p(\chi_{\rm p})\propto \frac{\chi_{\rm p}}{\sqrt{\tilde{a}^2 - \chi_{\rm p}^2}}$. Integrating, we find the cumulative distribution functions (CDF):
\begin{equation}\label{CDFeff}
N_{12}(\leq \chi_{\rm eff}) = 0.5 +0.75\frac{\chi_{\rm eff}}{\tilde{a}}~,
\end{equation}
and
\begin{equation}\label{CDFp}
N_{12}(\leq \chi_{\rm p}) = 1 - \sqrt{1 - \frac{\chi_{\rm p}^2}{\tilde{a}^2}}~.
\end{equation}

For 2G+2G mergers,
the probability density of $\chi_{\rm eff}$ is obtained  from the convolution 
of  two uniform distributions, which gives
$p({\rm \chi_{\rm eff}})\propto \left(\tilde{a}+ \chi_{\rm eff}\right)$ for $\chi_{\rm eff}\leq0$ and $p({\rm \chi_{\rm eff}})\propto\left(\tilde{a}-\chi_{\rm eff}\right)$ for $\chi_{\rm eff}>0$. The corresponding CDF is
\begin{equation}\label{22eff}
 N_{22}(\leq \chi_{\rm eff}) = \begin{cases} 
\frac{\chi_{\rm eff}}{\tilde{a}} + \frac{1}{2}{\chi_{\rm eff}^2\over \tilde{a}^2} +{1\over 2} & \text{if }  \chi_{\rm eff} \leq 0 \\
\frac{\chi_{\rm eff}}{\tilde{a}} - \frac{1}{2}{\chi_{\rm eff}^2\over \tilde{a}^2} +{1\over 2}
 & \text{if }\ \chi_{\rm eff}> 0 \ .
\end{cases} 
\end{equation}
Since $\theta_1$ and $\theta_2$
  are independent, the  distribution  of $\chi_{\rm p}$ 
    is obtained by the product of the distributions for each of the two angles:
  \begin{equation}\label{22p}
 N_{22}(\leq \chi_{\rm p}) = \left(1 - \sqrt{1 - \frac{\chi_{\rm p}^2}{\tilde{a}^2}}~\right)^2~ .
\end{equation}
{  We note that similar distributions for $\chi_{\rm eff}$ and   the
$\chi_{\rm eff}$ $vs$ $\chi_{\rm p}$ 
correlation were obtained by Ref.~\cite{2021PhRvD.104h4002B,2020PhRvD.102d3002B}.
}

We now asses the accuracy of the analytical model by comparing its predictions to the results of globular cluster models evolved using the fast cluster code {\tt cBHBd} \cite{2020MNRAS.492.2936A}.
Unless specified otherwise, our initial conditions and numerical modelling are the same as in~\cite{2023MNRAS.522..466A}. 
In all models considered here we adopt an initial cluster half-mass density
$\rho_{\rm h}=10^5M_\odot {\rm pc^{-3}}$ and the delayed supernova mechanism from \cite{Fryer2012}.

We report the results from four population models that differ by the  choice of the initial BH spin distributions and the maximum mass of 1G BHs, $m_{\rm cut}$.
In one model, the initial  BH spins are all set to zero and $m_{\rm cut}\simeq 70M_\odot$. 
In the other three models, we assume that $m_{\rm cut}\simeq 50M_\odot$ and that
the BH spin distribution follows a beta distribution with shape parameters
$(\alpha,\beta)=(2,18)$, $(2,5)$
and $(2,2)$; these distributions  peak at $ \simeq 0.06$, $0.2$, $0.5$ and their corresponding median values are $ \simeq 0.1$, 
$0.26$ and $0.5$, respectively.
 All but the last model (with the highest spins) give differential merger rates at primary masses above $m_{\rm cut}$ consistent with those measured from LIGO/Virgo data. 
Above this mass, the merger rate is dominated by 1G+2G mergers in all models (see the Supplementary Material; SM).

In Fig.~\ref{fig:simulation_cdf} we show the CDFs of $\chi_{\rm eff}$ and $\chi_{\rm p}$
for 1G+2G and 2G+2G mergers, as well as the analytical predictions based on equations~(\ref{CDFeff}),~ (\ref{CDFp}),~(\ref{22eff}) and (\ref{22p}), and the spin parameter distributions from the cluster Monte Carlo models of \cite{2019PhRvD.100d3027R}.
{  These distributions almost perfectly align with each other across  the entire range of parameter values. We note that while the differences near the tails of the distributions increase with initial BH spins, higher spins tend to reduce the merger rate above \( m_{\rm cut} \) due to increased BH ejection from their host clusters, so these models are statistically disfavored; on the other hand, the small deviations seen in the lower-spin models are unlikely to be discernible from the data.} Thus, if hierarchical mergers are the most common class of  merger above some threshold mass, our models predict a near-universal spin distribution, simply represented   by 
equations~(\ref{CDFeff}) and (\ref{CDFp}). In the following, we test this prediction
against GW data.

\begin{figure}
    \centering  
    \includegraphics[width=.45\textwidth]{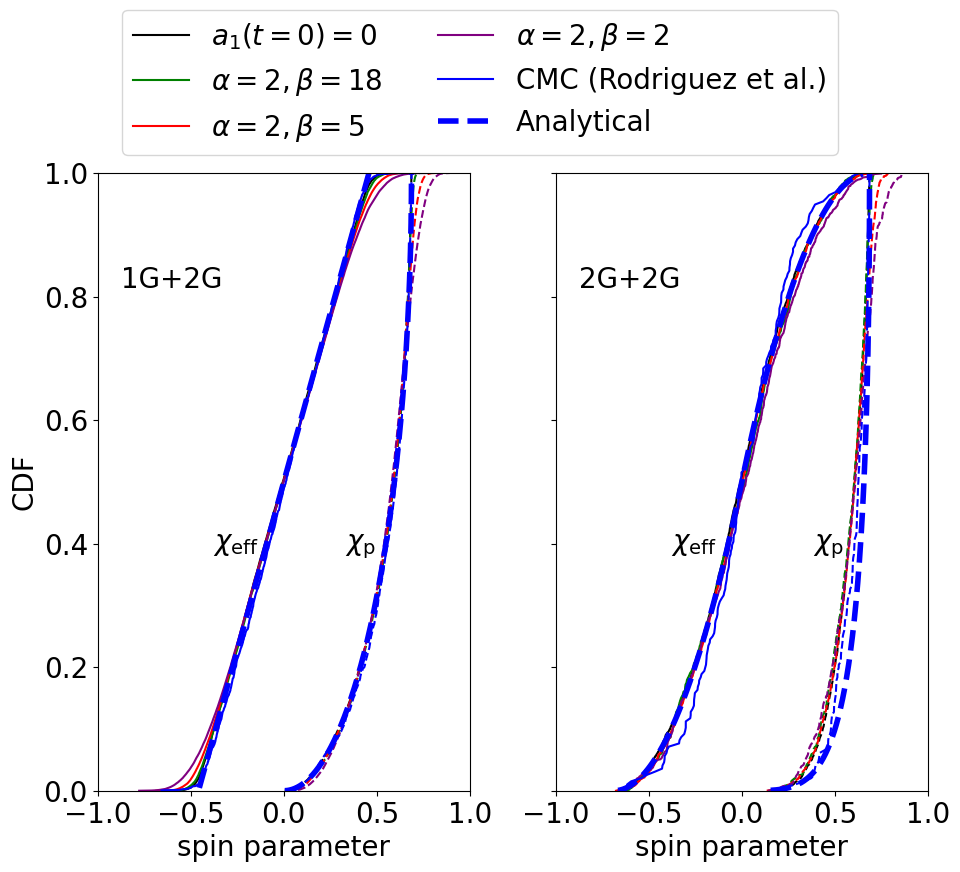}
    \caption{  CDF of $\chi_{\rm eff}$  and $ \chi_{\rm p}$ obtained from our cluster models and the expected CDFs based on our analytical approximations. The lines are hard to distinguish because they lie on top of each other, showing the independence of these distributions on model assumptions and initial conditions.
    }
    \label{fig:simulation_cdf}
\end{figure}

\noindent
{\it Bayesian inference.\ }Since the value of $\chi_{\rm p}$ is  less well-measured in GW observations than $\chi_{\rm eff}$ \cite{2019ApJ...882L..24A,2021arXiv211103606T}, we predominantly consider hierarchical inference of the effective spin distribution. 
For this study, we use the subset of BHs from GWTC-3 with false alarm rates below $1 \, {\rm yr}^{-1}${  , consistent with Ref.~\cite{LVKCollab2023}}. This results in a total of 69 binary BHs in our sample. 
{In parallel with} $\chi_{\rm eff}$, 
we hierarchically {fit} the distribution of  $q$,   $m_1$, and redshift $z$.
  We incorporate selection effects using the set of successfully
recovered binary BH injections provided by the LIGO-Virgo-KAGRA Collaboration and spanning their first three
observing runs \cite{LVKCollab2023,injections}.

\begin{figure}
 \hspace{-0.045\textwidth} 
    \centering  
\includegraphics[width=.7
\textwidth,angle =270]{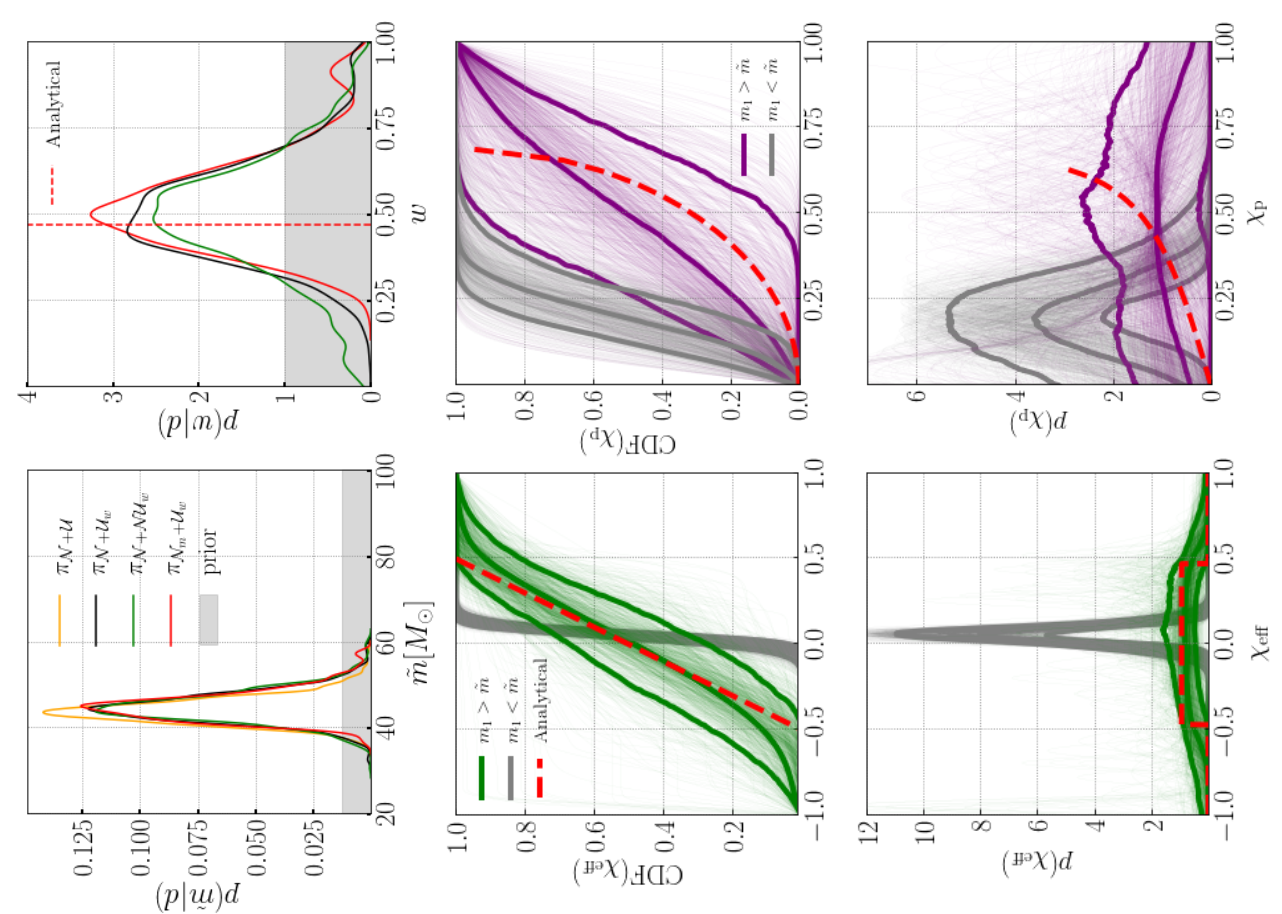}
    \caption{ Posteriors of $\tilde{m}$ (upper-left panel) and $w$ (upper-right panel) obtained under our models.   In the middle-left and bottom-left panels    
    we show
    the  distribution of $\chi_{\rm eff}$ for the $\pi_{\mathcal{N}+\mathcal{N}\mathcal{U}_w}$ model, for both the Gaussian component at $m_1\lesssim \tilde{m}$ and the  Uniform+Gaussian component at  $m_1\gtrsim \tilde{m}$.   
    We also show the  distribution of $\chi_{\rm p}$ obtained by using 
    equation~(\ref{eq:pi4}) in the middle-right and bottom-right panels.
    Thick lines are median, $10^{\rm th}$ and $90^{\rm th}$ quantiles, while  light lines are individual draws from the posterior.
Analytical lines are from equations~(\ref{CDFeff})
and ~(\ref{CDFp}).}
       \label{fig:posterior}
\end{figure}

We fit the $\chi_{\rm eff}$ distribution  to a
 mixture model comprising a Gaussian distribution, representing the bulk of the population at $m_1\lesssim \tilde{m}$, and a uniform distribution, representing  1G+2G hierarchical mergers at $m_1\gtrsim \tilde{m}$, where $\tilde{m}$ is the value of $m_1$ at which the transition between the Gaussian and uniform descriptions of $\chi_\mathrm{eff}$ occurs:
\begin{equation}
    \pi_{\mathcal{N}+\mathcal{U}}(\chi_\mathrm{eff}|m_1) =
        \begin{cases}
        \mathcal{N}(\chi_\mathrm{eff};\mu,\sigma) & (m_1<\tilde m) \\
        \mathcal{U}(\chi_\mathrm{eff};w=0.47) & (m_1 \geq \tilde m).
        \end{cases}
\label{eq:pi1}
\end{equation}
Here, $\mathcal{N}(\chi_\mathrm{eff};\mu,\sigma)$ denotes a normalized Gaussian distribution with mean $\mu$ and standard deviation $\sigma$ truncated within $[-1, 1]$, and $\mathcal{U}(\chi_\mathrm{eff}\,;w)$ is a uniform distribution defined over the range $|\chi_\mathrm{eff}|<w$.
We set $w = 0.47$, as predicted for hierarchical mergers.
We use broad uninformative priors; the prior on $\tilde{m}$ is uniform between $20$ and $100M_\odot$. For more details on the implementation of the models and prior assumptions
see the SM. 

We  consider  additional models with increased complexity.
Our second model,
$\pi_{\mathcal{N}+\mathcal{U}_w}(\chi_{\rm eff}|m_1)$,
is similar to $\pi_{\mathcal{N}+\mathcal{U}}$, but with $w$  now being a parameter that we infer from the data.
The third model is
\begin{equation}
\begin{aligned}
& \pi_{\mathcal{N}+\mathcal{N}\mathcal{U}_w}(\chi_\mathrm{eff}|m_1) \\
& =
        \begin{cases}
        \mathcal{N}(\chi_\mathrm{eff};\mu,\sigma) & (m_1<\tilde m) \\
      (1-\zeta)\mathcal{U}(\chi_\mathrm{eff};w) + \zeta \mathcal{N}_{\rm u}(\chi_\mathrm{eff};\mu_{\rm u}, \sigma_{\rm u}) & (m_1 \geq \tilde m),
        \end{cases}
\end{aligned}
\label{eq:pi3}
\end{equation}
where the mixing fraction $0\le\zeta\le 1$.
In this latter model, the potentially asymmetric distribution of $\chi_\mathrm{eff}$ above $\tilde{m}$ (if $\mu_\mathrm{u}$ is non-zero) allows us to assess how well the data support the theoretical expectation of the symmetry of $\chi_\mathrm{eff}$ around zero for $m_1 > \tilde{m}$ without this being enforced by the model.

If $\tilde{m}$ cannot be constrained or its posterior distribution rails against the limits of the prior, this would imply that a model with a separate spin distribution population above $\tilde{m}$ cannot be statistically distinguished from a Gaussian model applied across the entire mass range. 
A well-measured $\tilde{m}$, in turn, implies that there is a distinct mass threshold above which the population of binary BH mergers has a measurably distinct distribution of $\chi_\mathrm{eff}$. 
We measure $\tilde{m}=44^{+6}_{-4}M_\odot$ for the $\pi_{\mathcal{N}+\mathcal{U}}$ model (hereafter, reported measurements are median and $90\%$ credible interval).
The  posterior distributions of $\tilde{m}$ and $w$ are shown in Fig.~\ref{fig:posterior}. 
We infer $w=0.5^{+0.3}_{-0.2}$ and $0.5^{+0.3}_{-0.3}$ under the  $\pi_{\mathcal{N}+\mathcal{U}_w}$ and
the $\pi_{\mathcal{N}+\mathcal{NU}_w}$  models, respectively. While the recovered posteriors on $w$ are quite broad, they peak close to the expected value $w\simeq 0.47$  for hierarchical mergers.
We show in Fig.~\ref{fig:posterior} the  distribution of $\chi_{\rm eff}$
as inferred under $\pi_{\mathcal{N}+\mathcal{NU}_w}$.
Below $\tilde{m}$, the distribution is a narrow Gaussian with mean $\mu=0.05^{+0.03}_{-0.04}$; above $\tilde{m}$ the distribution is consistent with being symmetric around zero, and favors a large width as expected from a hierarchically formed population of mergers.
In the SM we show that our inference is unlikely to be affected by  
Monte Carlo uncertainty in the population likelihood estimator.

{  The precise measurement of $\tilde m$ implies that these models are strongly favored over models in which black holes of all masses share the same spin distribution.
More quantitatively, we compute the Bayes factors of models $\pi_{\mathcal{N}+\mathcal{U}}$, $\pi_{\mathcal{N}+\mathcal{U}_w}$, and $\pi_{\mathcal{N}+\mathcal{NU}_w}$ relative to a model where the entire population is represented by a single Gaussian in $\chi_{\rm eff}$ \cite[e.g.,][]{LVKCollab2023}.
We find $\log_{10} \mathcal{B} =4.7$, $4.5$ and $4.2$, respectively.
These results indicate strong evidence in favor of the hierarchical merger model over the Gaussian model above $\simeq 45M\odot$.}

Ref.~\cite{2022ApJ...932L..19B} identified a  broadening in the distribution of $\chi_\mathrm{eff}$ with increasing mass and/or redshift. {  Consistent results
were obtained by Ref.~\cite{2024PhRvD.109j3006H}
using flexible models that allow for a non-monotonic dependence of the mean and variance of the $\chi_{\rm eff}$ distribution on primary mass.}
The behavior we identify in this work -- the transition to a broad $\chi_\mathrm{eff}$ distribution above $\approx45\,M_\odot$ -- is likely responsible for this conclusion.
To verify this, we consider a fourth model, in which the fixed Gaussian spin distribution below $\tilde m$ is replaced with one whose mean and variance evolve linearly with primary mass, as in~\cite{2022ApJ...932L..19B}:
\begin{equation}\label{eq:pi1b}
\begin{aligned}
& \pi_{\mathcal{N}_m+\mathcal{U}_w}(\chi_\mathrm{eff}|m_1) =\\
 &   \begin{cases}
\mathcal{N}\left(\chi_\mathrm{eff};\mu_\chi(m_1),\sigma_\chi(m_1)\right) & (m_1<\tilde m)  \\
    \mathcal{U}(\chi_\mathrm{eff};w) & (m_1\geq \tilde m)
    \end{cases}
    \end{aligned}
\end{equation}
where $\mu_\chi(m_1)=\mu+\delta\mu\,(m_1/10\,M_\odot-1)$
and $\log \sigma_\chi(m_1)=\log \sigma+\delta \log \sigma(m_1/10\,M_\odot-1)$.
Under the  $\pi_{\mathcal{N}_m+\mathcal{U}_w}$ model 
we infer that mean and variance of effective spins below
$\tilde{m}$ are now consistent with no or mild change with
mass ($\delta \mu=0.00^{+0.02}_{-0.02}$, $\delta \log \sigma=-0.24^{+0.22}_{-0.07}$).
In fact, the data  prefer a reversal of the trend inferred in previous work~\cite{2022ApJ...932L..19B}, with the $\chi_\mathrm{eff}$ distribution slightly narrowing with mass below $\tilde m$.
Our interpretation is that the population above $\tilde{m}$ 
may drive the mass/redshift-dependent broadening reported in \cite{2022ApJ...932L..19B}.

Finally, we consider a  model where we also fit for the distribution of
 $\chi_{\rm p}$,
 which is
 represented
 by a mixture of two Gaussian distributions truncated within [0, 1], one below  and one above $\tilde{m}$:
\begin{equation}\label{eq:pi4}\small
\pi_{\chi_{\rm p}}(\chi_\mathrm{eff},\chi_\mathrm{p}|m_1) =
    \begin{cases}
 \mathcal{N}(\chi_{\rm eff};\mu,\sigma)\mathcal{N}(\chi_{\rm p};\mu_{\rm  p,l},\sigma_{\rm p,l})
     & (m_1<\tilde m)  \\
\mathcal{U}(\chi_{\rm eff};w)
   \mathcal{N}(\chi_{\rm p};\mu_{\rm  p,u},\sigma_{\rm p,u})
   & (m_1\geq \tilde m)
    \end{cases}
\end{equation}
Under this model, we infer $\tilde{m}=46^{+6}_{-4}M_\odot$
{  (see the SM for more details).
We
show 
the derived  distribution of $\chi_{\rm p}$  in  Fig.~\ref{fig:posterior}. The two distributions
 above and below $\tilde{m}$  can be nearly separated and the latter is broadly consistent with the
theoretical expectation. }
\\

\noindent
{\it Implications.}
We compare the predicted distribution of the effective spin parameters $\chi_{\rm eff}$ and $\chi_{\rm p}$ for hierarchical mergers to that of the population of observed BH binaries, finding evidence that the spin properties of observed binary BHs change at $\simeq 44M_\odot$. 
Above this mass, the $\chi_{\rm eff}$ distribution {  is consistent with the hypothesis of a (P)PISN gap in BH birth masses, which is repopulated by hierarchical mergers in dense clusters.
We infer that a fraction $p(m_1>\tilde{m})=0.009^{+0.01}_{-0.007}$ of binary BHs lie in the high-mass and isotropically spinning sub-population, with no appreciable variation across the different models. This can be interpreted as $\gtrsim1\%$ of binary BH mergers being of hierarchical origin---as these can also occur below $\tilde{m}$, albeit at a subdominant level---under our models.
}

 Sequential mergers in triples or quadruples \cite{2021ApJ...907L..19V,10.1093/mnras/stab178} are also possible, though the expected low merger rate makes this less likely \cite{2022MNRAS.516.1406S}. 
BHs in the (P)PISN mass gap can also arise from gas accretion and/or hierarchical mergers in AGN disks \cite{Bartos2016,Stone2016}, and stellar mergers in star clusters \cite{2020MNRAS.497.1043D,2021MNRAS.507.5132D}.
These latter scenarios are unlikely to give rise to the same  $\chi_{\rm eff}$ distributions as 
hierarchical BH mergers in clusters \cite{2019PhRvL.123r1101Y,PhysRevD.107.063007,2024A&A...685A..51V}, and are therefore disfavored, although not excluded, by our analysis as a primary formation mechanism. 

{  
Our findings align with previous work showing that the BH properties change above a certain mass, suggesting formation in  clusters for some fraction of the population \cite{2024arXiv240616844H,2024arXiv240601679P,2024arXiv240403166R,2023arXiv230401288G,2023arXiv230302973L,2022ApJ...935L..26F,2022arXiv220303651M}. 
Ref.~\cite{2023arXiv230302973L} and Ref.~\cite{2024arXiv240601679P} found evidence for a high-mass population consistent with hierarchical formation by studying the correlation between the  component spin distributions and mass. However, the component spin is not a parameter that is easily inferred  from the GW data, and its distribution  lacks robust constraints. Other parameters, e.g., mass, mass ratio, and merger rate, cannot be  robustly predicted from astrophysical models.  We have used  $\chi_{\rm eff}$ instead, which is both well constrained from the data,  and has a distribution that can be predicted from basic principles. This strengthens our conclusions and enables very stringent constraints on a hierarchical formation scenario with future data.
 Ref.~\cite{2022ApJ...935L..26F} also considered  the $\chi_{\rm eff}$ distribution and found  that the data allow for the presence of a high mass population consistent with hierarchical mergers, but do not require it. Here, we  showed that the data do require the presence of such a population.
}

The hierarchical formation hypothesis for high mass BHs makes other  predictions that will enable tighter constraints from future GW observations with a growing  population of BHs.
{  From the cluster population  models, we find that a $1\%$ fraction of hierarchical mergers in the (P)PISN mass gap implies that $\simeq 20\%$ of the BH binaries in the overall astrophysical population are formed in clusters.} Therefore, a significant fraction  of the detected binaries should present a residual eccentric signature~\cite{Antonini2014,2017arXiv171107452S,2018PhRvL.120o1101R}, of which traces are already claimed to exist in the data \cite{2022ApJ...940..171R,2024arXiv240414286G}. If mass-gap BHs form only  through hierarchical mergers in clusters, we should expect $\sim 5\%$ of mass-gap mergers to be measurably eccentric. Additionally, because BHs with mass above $\sim 90M_\odot$ can  only be formed from at least two previous mergers and these are rare, we might expect a drop/discontinuity in the  merger rate (and binary properties) near this mass.


\begin{acknowledgments}
\noindent{\it Acknowledgments.} FA is supported by the UK’s Science and Technology Facilities Council grant
ST/V005618/1. IMRS is supported by the Herchel Smith Postdoctoral Fellowship Fund. This material is based upon work supported by
NSF’s LIGO Laboratory which is a major facility fully funded by
the National Science Foundation, as well as the Science and Technology Facilities Council (STFC) of the United Kingdom, the Max-Planck-Society (MPS), and the State of Niedersachsen/Germany for
support of the construction of Advanced LIGO and construction and
operation of the GEO600 detector. Additional support for Advanced
LIGO was provided by the Australian Research Council. Virgo is
funded, through the European Gravitational Observatory (EGO), by
the French Centre National de Recherche Scientifique (CNRS), the
Italian Istituto Nazionale di Fisica Nucleare (INFN) and the Dutch
Nikhef, with contributions by institutions from Belgium, Germany,
Greece, Hungary, Ireland, Japan, Monaco, Poland, Portugal, Spain.
KAGRA is supported by Ministry of Education, Culture, Sports, Science and Technology (MEXT), Japan Society for the Promotion of
Science (JSPS) in Japan; National Research Foundation (NRF) and
Ministry of Science and ICT (MSIT) in Korea; Academia Sinica
(AS) and National Science and Technology Council (NSTC) in Taiwan. The authors are grateful for computational resources provided
by Cardiff University and supported by STFC grant
ST/V005618/1.

\noindent
{\it Main software:}
{\tt astropy} \cite{2013ascl.soft04002G};
{\tt bilby} \cite{2019ApJS..241...27A};
{\tt cBHBd} \cite{2020MNRAS.492.2936A};
{\tt jax} \cite{jax2018github};
{\tt numpy} \cite{2020Natur.585..357H};
{\tt numpyro} \cite{phan2019composable,bingham2019pyro};
{\tt scipy} \cite{2020SciPy-NMeth}.

\noindent{\it Data and code availability}. The cluster and hierarchical inference codes used to produce the results in this paper and the resulting data products are available under reasonable request.
\end{acknowledgments}

\bibliographystyle{apsrev4-1}
%

\clearpage

\setcounter{equation}{0}
\setcounter{figure}{0}
\setcounter{table}{0}

\renewcommand{\theequation}{S\arabic{equation}}
\renewcommand{\thefigure}{S\arabic{figure}}
\renewcommand{\thetable}{S\arabic{table}}

\setcounter{page}{1}
\appendix
\onecolumngrid
\section{Supplementary Material}\label{SM}
\date{\today}
\section{Globular cluster simulations}
We evolve a large number of globular cluster models using {\tt cBHBd}, which includes realistic BH initial mass functions \cite{Fryer2012} and a model for the cluster formation rate and evolution through redshift~\cite{2019MNRAS.482.4528E}. We consider 25 different values of metallicity between $0.01 Z_\odot$ and $Z_\odot$. These models follow the evolution of clusters and the formation of BH binary mergers through 3-body interactions. 
 The BHs are paired  using the paring functions derived in \cite{2023MNRAS.522..466A} using Heggie's theoretical interaction rate formulae \cite{1975MNRAS.173..729H}.
In all models considered here we adopt the delayed supernova mechanism \cite{Fryer2012} and an initial cluster half-mass density
$\rho_{\rm h}=10^5M_\odot {\rm pc^3}$.
{  The densities of Milky Way globular clusters reach up to approximately an order of magnitude lower than we have assumed \cite{PortegiesZwart2010}. However, it is important to note that the initial densities of these clusters were higher than their current densities. This aligns with our models, which evolve to significantly lower densities by the end of the simulations. Similarly, some nuclear clusters exhibit present-day densities near and above the assumed value \cite{2016MNRAS.457.2122G}.
When generating our model predictions, we take in account the uncertainties in the initial cluster mass function and  in the number density of clusters. This results in theoretical error bars in the merger rate estimates as in \cite{2020PhRvD.102l3016A}.}

\begin{figure*}
    \centering  
 \includegraphics[width=1.\textwidth]{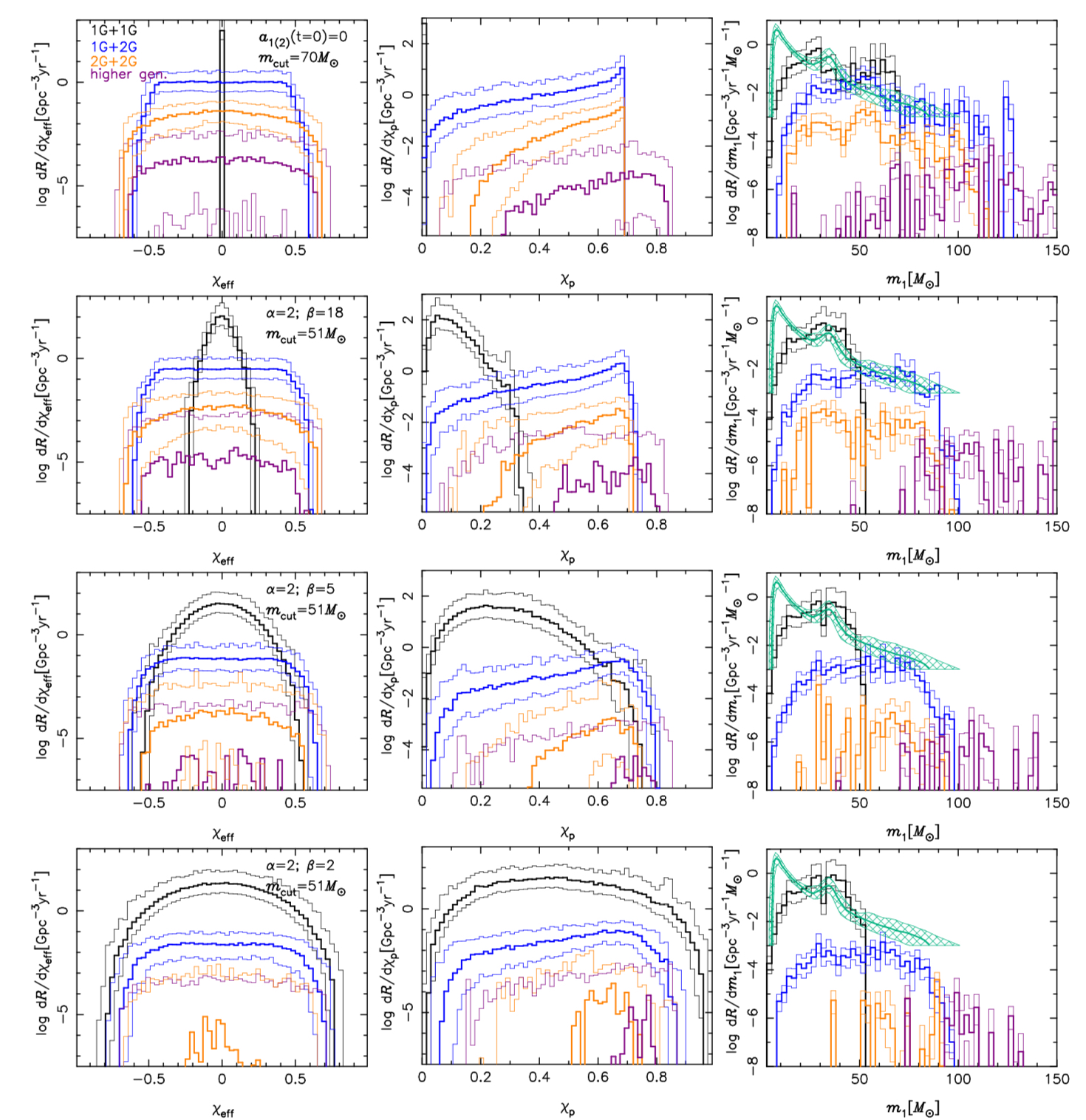} 
    \caption{ Differential merger rate of BH binaries formed in globular cluster simulations as a function of binary effective spin parameters $\chi_{\rm eff}$ (left panels), $\chi_{\rm p}$ (central panels) , and primary BH mass $m_1$ (right panels). 
The upper panels show the results for a model where the first generation BHs are formed with no spin, while in the other models     the BH birth spins follow a beta distribution with the $\alpha$ and $\beta$ parameters as shown.    
     The plot gives the median value of the merger rate in each bin, together with the 95 and 5 percentiles as thinner solid lines. 
     {  These errors   reflect uncertainties in the initial cluster mass function and  number density of clusters.}
     In the right panel, the green curve and hatched regions show the  mean merger rate as inferred from the GW data and the corresponding confidence intervals as reported by \cite{LVKCollab2023} under the \textsc{PowerLaw + Peak} model.          
    }
    \label{fig:simulation_pdfs}
\end{figure*}

We report the results from four models that differ by the  choice of the initial BH spin distribution. In one model the initial BH spins are all set to zero. Moreover, we do not include any prescription for pair-instability processes. Nevertheless, we find that no 1G BH is formed with a mass above $m_{\rm cut}\simeq70 M_\odot$, implying a mass truncation set by the initial upper limit on the stellar mass function (here at $130M_\odot$) and by stellar wind mass-loss prior to the formation of the BHs.
In a second model, we assume that the initial BH spin distribution follows a beta distribution with $\alpha=2$ and
$\beta=18$, which peaks at $a \simeq 0.06$ and has a median at $\simeq 0.1$. In a third model, the initial BH spin distribution follows a beta distribution with $\alpha=2$ and
$\beta=5$, which  peaks at $a \simeq 0.2$ and has a median at $\simeq 0.26$. In a fourth model, we use a beta distribution with $\alpha=2$ and
$\beta=5$, which  peaks at $0.5$ and has a median at $0.5$.
In these latter three models, $m_{\rm cut}\simeq 51 M_\odot$, as determined by the adopted prescriptions for pair instability 
\cite[taken from ][]{2017MNRAS.470.4739S}.
All models result in a sharp truncation of the initial BH mass function at $m_{\rm cut}$. We note that certain high-mass star properties not included in our models, such as rotation or chemical composition, could result in the formation of BHs exceeding this limit. Therefore, the existence of a strict mass cut-off should be regarded as an assumption.

In Fig.~\ref{fig:simulation_pdfs} we show the differential merger rate of BH binaries as a function of $\chi_{\rm eff}$, $\chi_{\rm p}$ and $m_1$. We separately plot the populations of 1G+1G mergers, 1G+2G mergers, 2G+2G mergers, and the remaining mergers that involve higher generation BHs, the majority of which are 3G+1G and 3G+2G mergers. The figure shows that 
the simulated populations cannot explain the
peak at $10M_\odot$ inferred from the GW data under the \textsc{PowerLaw + Peak} model from GWTC-3 \citep{LVKCollab2023}, but can account for all mergers with primary mass 
$m_1\gtrsim 30M_\odot$. 
{  The lower edge of the mass gap can be identified 
as the upper edge of the $m_1$ distribution where the 1G+1G merger rate drops to zero, and naturally coincides with the initial upper limit of the initial BH mass function that is set by our stellar evolution recipes.
In all models the merger rate above this limit is dominated by 1G+2G mergers.}   As expected based on our theoretical considerations, the left panel of Fig.~\ref{fig:simulation_pdfs} shows that  their  $\chi_{\rm eff}$ distribution is nearly uniform (i.e., flat)  within $|\chi_{\rm eff}|< 0.5$; the distribution of $\chi_{\rm p}$ is peaked at $\simeq 0.7$; and as the 1G BH spin increases, the overall contribution of 1G+2G binaries to the merger rate decreases. 
Naturally, 1G+2G as well as 2G+2G mergers have a null contribution to the merger rate above $m_1>2m_{\rm cut}$. In this range of masses, the merging binaries are 3G+1G, 3G+2G and 4G+1G.

{  We note that
Ref.~\cite{2021ApJ...915L..35K} used a
phenomenological population model and the data from 
the second LIGO–Virgo Gravitational-Wave
Transient Catalog to infer the median relative merger rates of 1G+2G and
2G+2G to 1G+1G mergers to be
$\sim 10^{-2}$ and $\sim 10^{-5}$. These numbers are broadly consistent with the number ratio between the two populations found in our models.}

\begin{figure*}
    \centering  
    \includegraphics[width=1.\textwidth]{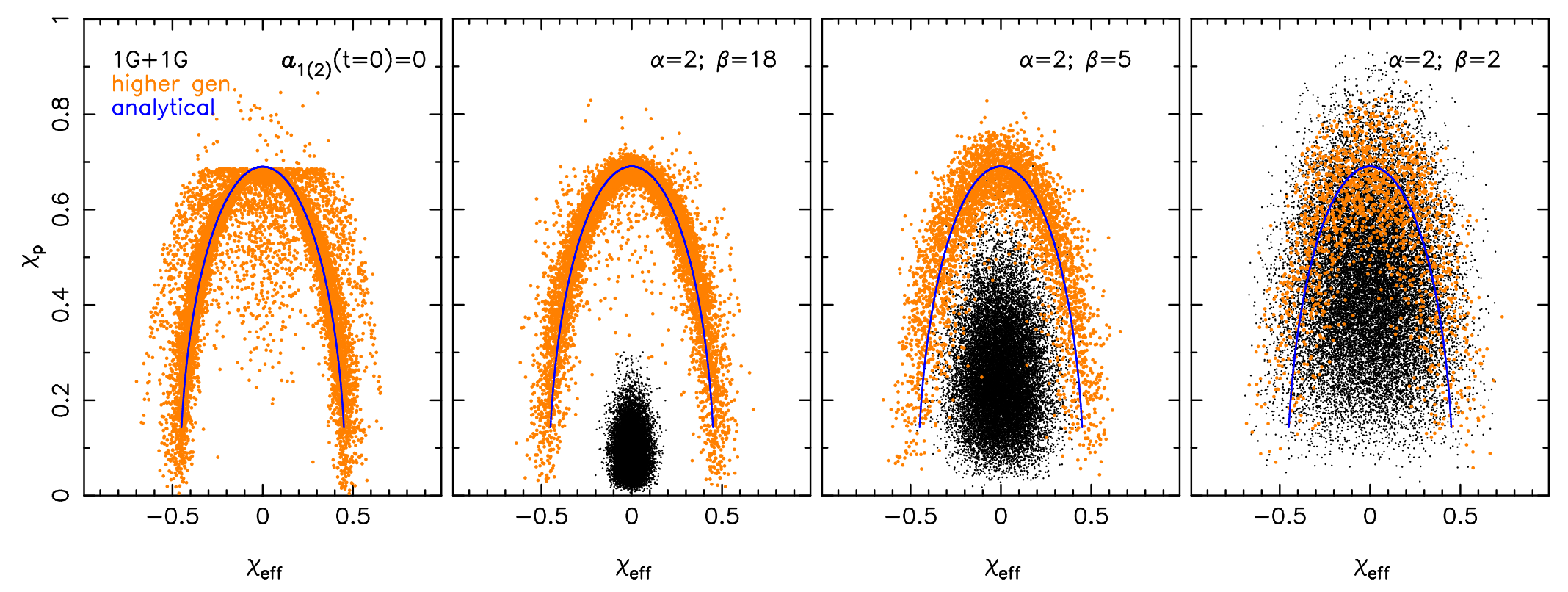} 
    \caption{
Values of $\chi_{\rm eff}$ and $\chi_{\rm p}$ for
a fraction  of randomly selected binary BH mergers formed in our cluster models. The blue line shows the mean correlation predicted by our simple analytical model.}
    \label{fig:simulation_scatter}
\end{figure*}

Fig.~\ref{fig:simulation_scatter}  shows  $\chi_{\rm eff}$ ~ vs  $\chi_{\rm p}$ for a fraction of the binaries that merge in each model. The mean correlation between these two parameters is  well described by our simple model. The model with 
$(\alpha,\beta)=(2,2)$ demonstrates significant overlap in the spin parameter distributions for 1G+1G and 1G+2G mergers, while still showing a substantial merger rate contribution from 1G+2G mergers. Thus, a confidently measurable value of $\chi_{\rm p}$ alone cannot be used to identify a single detected binary as hierarchically formed if BH natal spins are high. On the other hand, the population properties can be clearly used to put a hierarchical formation scenario to the test, as we do in this work.

\begin{table}
\begin{center}
 \renewcommand{\arraystretch}{1.05}
 \renewcommand{\tabcolsep}{0.25cm}
 \begin{tabular}{c | l | l l l l l}
 \hline
 \hline
 Parameter 
    & Prior 
    & $\mathcal{N}+\mathcal{U}$
    & $\mathcal{N}+\mathcal{U}_w$
    & $\mathcal{N}+\mathcal{N}\mathcal{U}_w$ 
    & $\mathcal{N}_m+\mathcal{U}_w$
    & $\chi_\mathrm{p}$ \\
 \hline
  $\tilde{m}$
    & $\mathcal{U}(20,100)$
    & \checkmark
    & \checkmark 
    & \checkmark 
    & \checkmark 
    & \checkmark \\
$w$ 
    & $\mathcal{U}(0,1)$
    & --
    & \checkmark
    & \checkmark 
    & \checkmark 
    & \checkmark \\
$\mu$   
    & $\mathcal{U}(-1,1)$ 
    & \checkmark
    & \checkmark
    & \checkmark
    & \checkmark
    & \checkmark
    \\
$\zeta$
    & $\mathcal{U}(0,1)$
    & --
    & --
    & \checkmark
    & --
    & --
    \\
$\mu_{\rm u}$  
    & $\mathcal{U}(-1,1)$ 
    & --
    & --
    & \checkmark
    & --
    & --
    \\
$\log_{10}\sigma$   
    & $\mathcal{U}(-1.5,0)$ 
    & \checkmark
    & \checkmark
    & \checkmark
    & \checkmark
    & \checkmark
    \\
$\log_{10}\sigma_{\rm u}$  
    & $\mathcal{U}(-1.5,0)$ 
    & --
    & --
    & \checkmark
    & --
    & --
    \\
$\delta \mu$ 
    & $\mathcal{U}(-2.5,1)$ 
    & --
    & --
    & --
    & \checkmark
    & --
    \\
$\delta \log\sigma$
    & $\mathcal{U}(-2,1.5)$
    & --
    & --
    & --
    & \checkmark
    & --
    \\
$\mu_{\rm p, u}$ and $\mu_{\rm p, l}$
    & $\mathcal{U}(0.02,1)$
    & --
    & --
    & --
    & --
    & \checkmark
    \\
$\log_{10}\sigma_{\rm p,u}$ and $\log_{10}\sigma_{\rm p, l}$ 
    & $\mathcal{U}(-1.2,0)$
    & --
    & --
    & --
    & --
    & \checkmark
    \\
 \hline
 \hline
\end{tabular}
\caption{
Priors adopted on the hyperparameters describing the effective spin distribution of binary BHs.
For each parameter, we additionally indicate the model(s) in which the parameter appears.
}
\label{tab:spin-priors}
\end{center}
\end{table}

\begin{table}
\begin{center}
 \renewcommand{\arraystretch}{1.05}
 \begin{tabular}{c l l}
 \hline
 \hline
 Parameter & Prior & Defined in \\
 \hline
$\beta_q$ & $\mathcal{N}(0,3)$ & equation~\eqref{eq:pm2} \\
$\kappa$ & $\mathcal{N}(0,6)$ & equation~\eqref{eq:pz} \\
$\mu_m$ & $\mathcal{U}(50\,M_\odot,100\,M_\odot)$ & equation~\eqref{eq:pm} \\
$\sigma_m$ & $\mathcal{U}(2\,M_\odot,15\,M_\odot)$ & equation~\eqref{eq:pm} \\
$f_p$ & $\mathcal{U}(0,1)$ & equation~\eqref{eq:pm} \\
$\lambda$ & $\mathcal{N}(-2,3)$ & equation~\eqref{eq:pm} \\
$ m_{\rm max}$ & $\mathcal{U}(60\,M_\odot,100\,M_\odot)$ & equation~\eqref{eq:pm} \\
$ m_{\rm min}$ & $\mathcal{U}(5\,M_\odot,15\,M_\odot)$ & equation~\eqref{eq:pm} \\
$ \log_{10} dm_{\rm max}/\,M_\odot$ & $\mathcal{U}(0.5,1.5)$ & equation~\eqref{eq:pm} \\
$ \log_{10} dm_{\rm min}/\,M_\odot$ & $\mathcal{U}(-1,1)$ & equation~\eqref{eq:pm} \\
 \hline
 \hline
\end{tabular}
\caption{
Priors adopted for the hyperparameters with which we describe the primary mass, mass ratio, and redshift distributions of the binary BH population.
}
\label{tab:priors}
\end{center}
\end{table}



\subsection{Population models, data and hierarchical inference}

We perform hierarchical Bayesian inferences to fit the data
of the observed events  with a given population model.
Here, we summarise the main ingredients of the models, the data we use, and our inference analysis method.

\begin{figure*}
    \centering  
    \includegraphics[width=1.\textwidth]{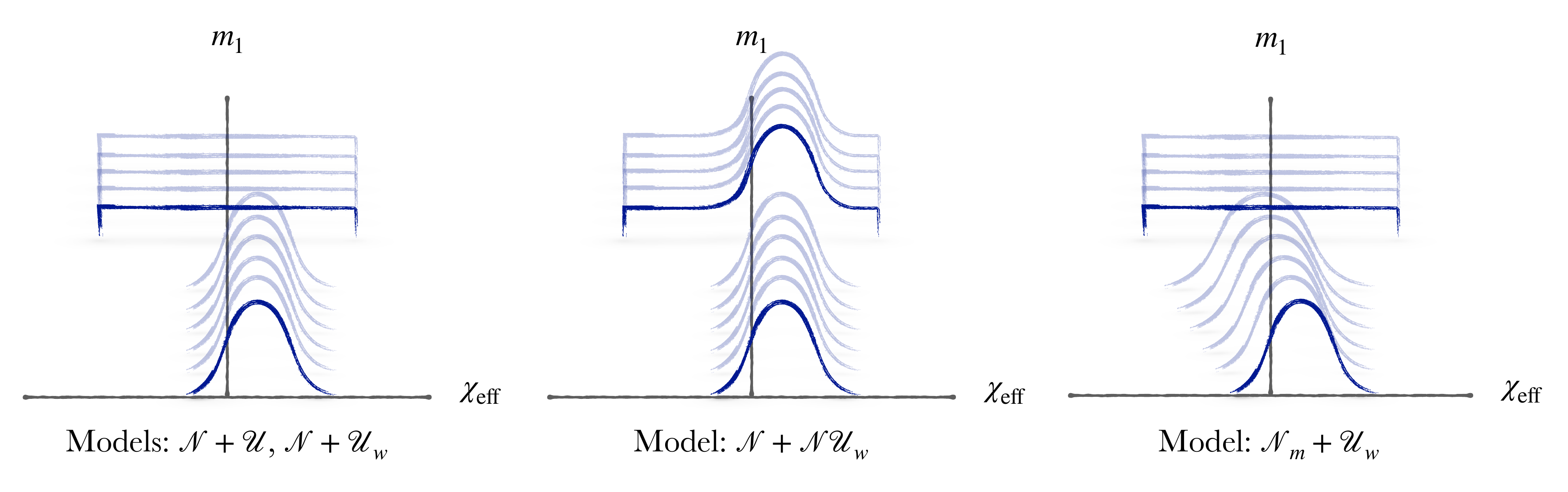} 
    \caption{
    Cartoon illustrating the main models examined in the main text. {  Blue lines show a  representation of the parametric models used in our inference analysis.}
    }
    \label{fig:cartoon}
\end{figure*}

As described in the main text, we investigate several different models for the $\chi_\mathrm{eff}$ distribution of binary BHs as a function of their $m_1$.
These are sketched in Fig.~\ref{fig:cartoon}.
Illustrated in the left-hand subplot, models $\pi_{\mathcal{N}+\mathcal{U}}$ and $\pi_{\mathcal{N}+\mathcal{U}_w}$ (see equation~\eqref{eq:pi1}) assume that the effective spin distributions transitions from a Gaussian to a flat uniform distribution above some threshold mass $\tilde{m}$.
Shown in the middle subplot, model $\pi_{\mathcal{N}+\mathcal{N}\mathcal{U}_w}$ (see equation~\eqref{eq:pi3}) instead assumes that the Gaussian spin distribution persists across both low- and high-mass systems, but includes a uniform component above $\tilde{m}$.
Finally, in the right-hand panel, model $\pi_{\mathcal{N}_m+\mathcal{U}_w}$ (see equation~\eqref{eq:pi1b}) adopts a low-mass Gaussian whose mean and log-standard deviation vary linearly as a function of mass.
The priors on the hyperparameters defining each model are listed in Table~\ref{tab:spin-priors}.

We perform hierarchical inference on the binary BH population using the Hamiltonian Monte Carlo sampling algorithm implemented \texttt{numpyro}, a probabilistic programming library based on \texttt{jax}.
Hamiltonian Monte Carlo methods require a likelihood that is a differentiable function of the population hyperparameters.
However, the piecewise equations that define our spin models (e.g. equation~\eqref{eq:pi1}) are not differentiable; the discontinuities at $m_1 = \tilde m$ cause the likelihood to itself discontinuously as a function of $\tilde m$.
To remedy this, in practice we implement these piecewise models as sharp but continuous transitions in the effective spin distributions above and below $\tilde m$.
Our baseline model $\pi_{\mathcal{N}+\mathcal{U}}$ defined in equation~\eqref{eq:pi1}, for example, is approximated as
    \begin{equation}
    \pi_{\mathcal{N}+\mathcal{U}}(\chi_\mathrm{eff}|m_1) = 
        \left[1-\eta(m_1)\right]\,\mathcal{N}(\chi_\mathrm{eff};\mu,\sigma)
        + \eta(m_1)\,\mathcal{U}(\chi_\mathrm{eff};w=0.47),
    \end{equation}
where 
    \begin{equation}
    \eta(m_1)
    = \left[{1 + \exp\left(-\frac{(m_1 - \tilde{m})}{3\,M_\odot}\right)}\right]^{-1}
    \end{equation}
is a logistic function that is approximately equal to zero below $\tilde m$ and unity above.
Other spin models are implemented analogously.
Similarly, a true truncated uniform distribution at high masses would cause the likelihood to change discontinuously with $w$.
When implementing $\mathcal{U}(\chi_\mathrm{eff};w)$ we therefore do not use an infinitely sharp truncation at the boundaries but instead exponentially suppress the distribution beyond $|\chi_\mathrm{eff}| = w$:
    \begin{equation}
    \mathcal{U}(\chi_\mathrm{eff};w) \propto
        \begin{cases}
        1 & (|\chi_\mathrm{eff}|\leq w) \\
        \mathrm{exp}\left[-\frac{(|\chi_\mathrm{eff}|-w)^2}{2 \left(0.1\right)^2}\right] & (|\chi_\mathrm{eff}| > w).
        \end{cases}
    \end{equation}
Our conclusions do not depend on the precise scales over which the above smoothing takes place.

Alongside the effective spin distribution, we hierarchically measure the
distribution of binary BH primary masses $m_1$, mass ratios $q$, and redshifts $z$.
We model the $m_1$ distribution as a mixture between a power law and a Gaussian, with exponential tapering functions at high and low masses:
\begin{eqnarray}
\label{eq:pm}
p(m_1) \propto
    T_{\rm l}(m_1)
    T_{\rm h}(m_1)
 \Big[(1- f_p) P(m_1) + f_p \,\mathcal{N}(m_1) \Big].
\end{eqnarray}
Here, $P(m_1) \propto m_1^\lambda$ and $\mathcal{N}(m_1) \propto \exp\left[-\frac{(m_1-\mu_m)^2}{2\sigma_m^2}\right]$ are power-law and Gaussian distributions, each normalized over the interval $2\,M_\odot\leq m_1\leq 100\,M_\odot$.
The tapering functions are defined as
    \begin{equation}
    T_{\rm l}(m_1) =\begin{cases}
        \exp\left[-\frac{(m_1 - m_{\text{min}})^2}{2 d m_{\text{min}}^2}\right] & (m_1<m_{\rm min}) \\
        1 & (m_1\ge m_{\rm min})
        \end{cases}
    \end{equation}
and
    \begin{equation}
    T_{\rm h}(m_1) = \begin{cases}
        1 & (m_1 \leq m_\mathrm{max}) \\
        \exp\left[-\frac{(m_1 - m_{\text{max}})^2}{2 d m_{\text{max}}^2}\right] & (m_1 > m_{\text{max}})
    \end{cases}
    \end{equation}
We assume that the secondary mass $m_2$ distribution follows~\cite{2022ApJ...937L..13C}
    \begin{equation}
    \label{eq:pm2}
    \centering
    p(m_2|m_1)\propto m_2^{\beta_q} \quad (2\,M_\odot \leq m_2\leq m_1)\ .
    \end{equation}
Finally, we assume a distribution of $z$ that is proportional to the differential comoving volume $dV_c/dz$, with a possible evolution in the merger rate towards higher $z$~ \cite{Fishbach_2018,Callister_2020}
 \begin{equation}
 \label{eq:pz}
p(z)\propto {1\over 1+z} {dV_c\over dz} (1+z)^{\kappa}\ .
 \end{equation}
The priors placed on the hyperparameters governing the $m_1$, $q$, and $z$ distributions are listed in Table~\ref{tab:priors}.
       
We perform our inference using the subset of binary BH events from GWTC-3 with false alarm rates below $1\,\mathrm{yr}^{-1}$.
We exclude GW170817, GW190425, GW190426, GW190814, GW190917, GW200105, GW200115~\cite{2021PhRvX..11b1053A,2021arXiv211103606T}, as they
have at least one component with a mass $< 3\,M_\odot$ 
{  that are most likely neutron stars \cite{LVKCollab2023}.}
This leaves a total of $N_{\rm det} = 69$ binary black holes in our sample. 
We use parameter estimation samples made publicly available through the \ \href{https://www.gw-openscience.org/}{Gravitational-Wave Open Science Center}.
For events first published in \href{https://dcc.ligo.org/LIGO-P1800370/public}{GWTC-1}~\cite{2019PhRvX...9c1040A}, we use the “\texttt{Overall\_posterior}” parameter estimation samples.
For events first published in \href{https://dcc.ligo.org/LIGO-P2000223/public}{GWTC-2}~\cite{2021PhRvX..11b1053A} and 
\href{https://zenodo.org/record/5117703}{GWTC-2.1}~\cite{2024PhRvD.109b2001A}, we adopt the “\texttt{PrecessingSpinIMRHM}” samples, and for  events in GWTC-3~\cite{2021arXiv211103606T}, we use the “\texttt{C01:Mixed}” samples~\footnote{GWTC-3 samples available at https://zenodo.org/record/5546663}.
These selections correspond to a union of samples obtained with different waveform families.
All samples account for spin precession effects, while the \texttt{PrecessingSpinIMRHM} and \texttt{C01:Mixed} samples from GWTC-2, GWTC-2.1, and GWTC-3 additionally include the effects of higher order modes (parameter estimation incorporating higher order modes was not available in GWTC-1).
We assess the detection efficiency using the set of successfully recovered binary BH injections, provided by the LIGO-Virgo-KAGRA collaborations, spanning their first three observing runs~\cite{injections}.

We perform our analysis using standard hierarchical inference.
Let $p(\theta_i|d_i)$ be posteriors on the individual parameters $\theta_i$ (e.g. component masses, redshift, etc.) of each gravitational-wave event, conditioned on its observed data $d_i$.
The corresponding posterior on the population parameters $\Lambda$ is~\cite[e.g.,][]{
Fishbach_2018,2019MNRAS.486.1086M,2022ApJ...937L..13C}
    \begin{equation}
    p(\Lambda \,|\, \{d_i\})
        \propto p(\Lambda)\, \xi^{-N_{\rm det}}(\Lambda)
        \prod_{i=1}^{N_{\rm det}}
        \int d\theta_i\, p(\theta_i|d_i)
        \frac{p(\theta_i|\Lambda)}{p_\mathrm{pe}(\theta_i)}\ ,
    \label{eq:likelihood-integral}
    \end{equation}
where $p_\mathrm{pe}(\theta_i)$ is the  prior adopted for purposes of parameter estimation and  $p(\Lambda)$ is our prior on the population-level parameters.
We use the priors listed in Tables~\ref{tab:spin-priors} and~\ref{tab:priors}.
We replace integration over $p(\theta_i|d_i)$ with an ensemble average taken over the posterior samples associated with each event:
    \begin{equation}
    p(\Lambda \,|\, \{d_i\})
        \propto p(\Lambda)\,\xi^{-N_{\rm det}}(\Lambda)
        \prod_{i=1}^{N_{\rm det}}
        \bigg\langle
        \frac{p(\theta_i|\Lambda)}{p_\mathrm{pe}(\theta_i)}
        \bigg\rangle.
    \label{eq:likelihood-sum}
    \end{equation}
{  Given a number $N_{\rm inj}$ of injected
signals drawn from some reference distribution $p_\mathrm{inj}(\rm \theta_i)$
The detection efficiency $\xi(\Lambda)$ quantifies the total fraction of events that we expect to pass our detection criteria:
\begin{equation}\label{csi}
\xi(\Lambda)={1\over N_{\rm inj}}
\sum^{N_{\rm found}}_{i=1}\frac{p(\theta_i|\Lambda)}{p_\mathrm{inj}(\rm \theta_i)},
\end{equation}
summing over the $N_\mathrm{found}$ injections that pass our detection
criteria}, and where $N_{\rm inj}$ is the total number of injections (including those that are not recovered).
We estimate $\xi(\Lambda)$ using the injection campaign reported in \cite{2024PhRvD.109b2001A,injections}, selecting successfully found injections (with recovered false alarm rates below $1$~yr$^{-1}$ in at least one pipeline) and reweighting to the proposed population $\Lambda$ 
as in \cite{2022ApJ...937L..13C}.
We sample over equation~\eqref{eq:likelihood-sum} using the \textsc{numpyro}'s~\cite{phan2019composable,bingham2019pyro} implementation of the ``No U-Turn'' Hamiltonian Monte Carlo algorithm~\cite{2011arXiv1111.4246H}.

\begin{figure*}
    \centering  
    \includegraphics[width=1.\textwidth]{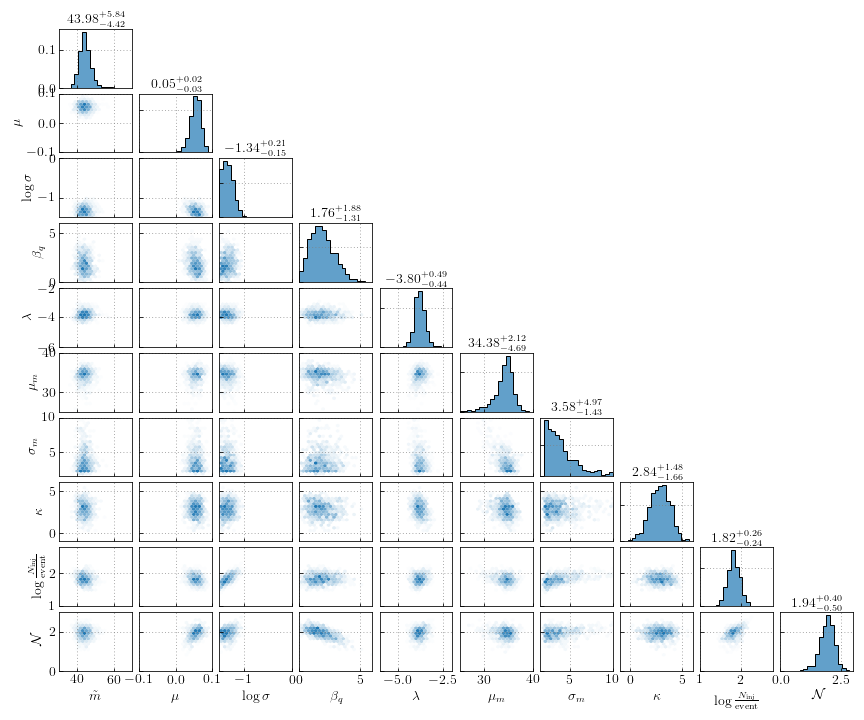}
    \caption{Posteriors on the  parameters that govern the hierarchical model where
    the $\chi_{\rm eff}$ distribution is
    given by equation~(\ref{eq:pi1}): a Gaussian below $\tilde{m}$ and
    a Uniform distribution with half width $w=0.47$ above $\tilde{m}$. Here $ \log$ indicates $\log_{10}$ and $N_{\rm inj}/event$ gives the total number of injection divided by the number of  detections.
    }
    \label{fig:post1}
\end{figure*}

\begin{figure*}
    \centering  
    \includegraphics[width=1.\textwidth]{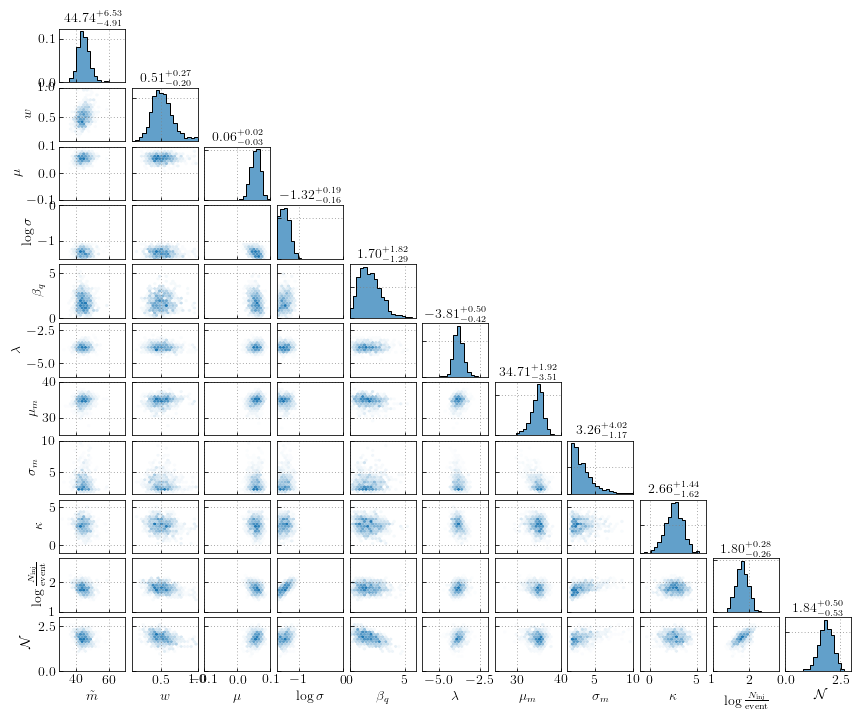}
    \caption{Posteriors on the parameters that govern the hierarchical model where
    the $\chi_{\rm eff}$ distribution is
    represented by 
    a fixed Gaussian 
    below $\tilde{m}$
    and a Uniform distribution with variable half width, $w$, above $\tilde{m}$. Here $ \log$ indicates $\log_{10}$ and $N_{\rm inj}/event$ gives the total number of injection divided by the number of  detections.
    }
    \label{fig:post2}
\end{figure*}

\begin{figure*}
    \centering  
    \includegraphics[width=1.\textwidth]{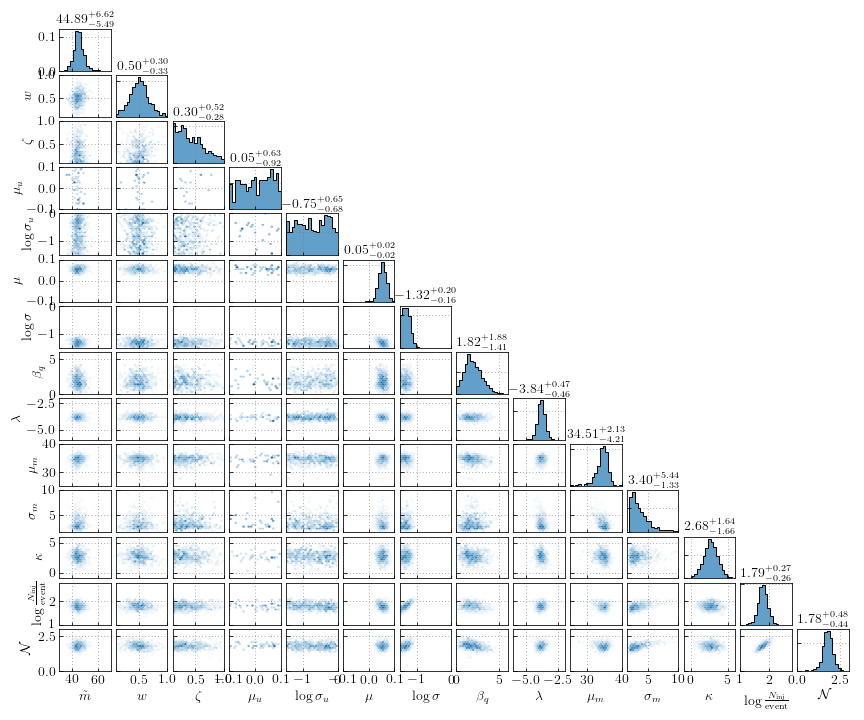}
    \caption{Posteriors on the  parameters that govern the hierarchical model where
    the $\chi_{\rm eff}$ distribution is
    represented by equation~(\ref{eq:pi3}):
     a fixed Gaussian 
    below $\tilde{m}$
    and a Uniform distribution with variable half width, $w$, plus a Gaussian above $\tilde{m}$. Here $ \log$ indicates $\log_{10}$ and $N_{\rm inj}/event$ gives the total number of injection divided by the number of  detections.
    }
    \label{fig:post3}
\end{figure*}

\begin{figure*}
    \centering  
    \includegraphics[width=1.\textwidth]{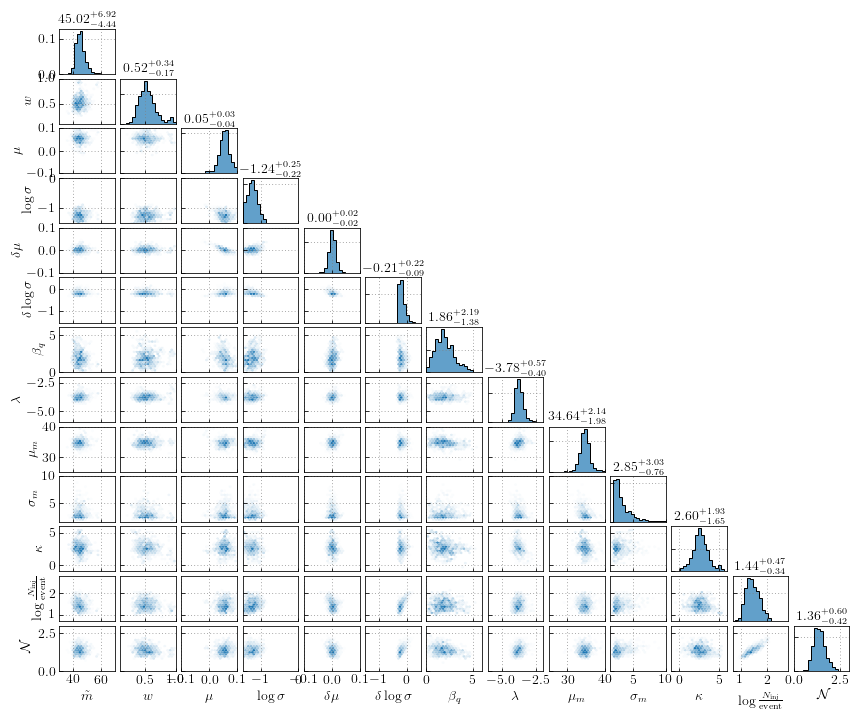}
    \caption{Posteriors on the  parameters that govern the hierarchical model where
    the $\chi_{\rm eff}$ distribution is
    represented by equation~(\ref{eq:pi1b}):
    a Gaussian with mass dependent mean and variance below 
    $\tilde{m}$ and
    a fixed Uniform distribution with half width $w$ above $\tilde{m}$. Here $ \log$ indicates $\log_{10}$ and $N_{\rm inj}/event$ gives the total number of injection divided by the number of  detections.
    }
    \label{fig:post1b}
\end{figure*}

The full posterior distributions on the key parameters that enter in our population analysis are given in Figs.~\ref{fig:post1},
 \ref{fig:post2},
 \ref{fig:post3}, and \ref{fig:post1b}.
{  In particular, we note that although parameter $\zeta$ in $\pi_{\mathcal{N}+\mathcal{NU}_w}$ favors a value of zero (i.e. a full transition to a uniform $\chi_{\rm eff}$ distribution above $\tilde m$), it has large uncertainties, being constrained to lie in the range of 0.02--0.82 in $90\%$ credible intervals.
This might initially appear to reduce the robustness of our conclusions, implying that the data are unable to distinguish with confidence a uniform from a Gaussian distribution above $\tilde{m}$.
This is not surprising, however, and suggests that our inference is not particularly sensitive to how the high mass population is parametrized.
In fact, we consider a  model where the high mass population is represented by a Gaussian (by setting  $\zeta=1$), and find a consistent  distribution  of $\chi_{\rm eff}$ and  $\tilde{m}=45^{+5}_{-4}M_\odot$.
Finally, we note that in the $\pi_{\mathcal{N}+\mathcal{NU}_w}$ model the parameters of the Gaussian are essentially unconstrained, which means that this distribution is probably not required.
}

{ 
In the main text, we quote Bayes factors between our models and one in which the binary BH effective spin distribution is modeled as a single Gaussian, $\pi_{\mathcal{N}}$, with no distinct high-mass population.
Bayes factors are obtained by sampling mixture models of the form $\pi_{\rm mix}(\chi_{\rm eff}|m_1) = \xi \,\pi_\mathcal{A}(\chi_{\rm eff}|m_1) + (1-\xi)\, \pi_\mathcal{N}(\chi_{\rm eff}|m_1)$, where $\pi_\mathcal{A}(\chi_{\rm eff}|m_1)$ is one of our models under investigation.
The quantity $\xi$ is not continuous, but is a \textit{categorical} variable that can be either $0$ or $1$, corresponding to populations described entirely by $\pi_{\mathcal{N}}$ or $\pi_{\mathcal{A}}$, respectively.
We sample this likelihood using the \texttt{DiscreteHMCGibbs} sampler implemented in \texttt{numpyro}; this performs Hamiltonian Monte Carlo over the continuous parameters defining models $\pi_{\mathcal{N}}$ and $\pi_{\mathcal{A}}$ while sampling the categorical variable $\xi$ using Metropolis-Hastings.
The Bayes factor between models $\pi_{\mathcal{A}}$ and $\pi_{\mathcal{N}}$ is then given by the ratio of posterior probabilities $p(\xi=1|\{d_i\})/p(\xi=0|\{d_i\})$.
}

We also perform several leave-one-out analyses to determine whether our results are driven primarily by a small number of unusual events.
In particular, the events GW170729 and GW190517 could conceivably be driving the preference for a broad spin distribution at high masses; both events have confidently large spins ($\chi_\mathrm{eff} = 0.36^{+0.21}_{-0.25}$ and $0.52^{+0.20}_{-0.19}$, respectively, under default priors) and have primary masses ($50^{+16}_{-10}$ and $36^{+12}_{-8}\,M_\odot$) near our inferred values of $w$ and $\tilde m$.
We repeat our analysis under the $\pi_{\mathcal{N}+\mathcal{U}_w}$ model excluding one or both of GW170729 and GW190517.
In all cases, results remain consistent with those shown in Fig.~\ref{fig:posterior}, increasing our confidence that we are identifying a feature inherent in the broader black hole population.

Finally, in order to identify possible issues due to  finite sampling effects when estimating Eq.~\ref{eq:likelihood-sum}, we track the number of ``effective samples'' $N_{\mathrm{eff}}$ informing the Monte Carlo estimates of the likelihood for every event.
Given a set of $N_i$ posterior samples $\{\lambda_{i,j}\}_{j=1}^{N_i}$ for each event $i$, the $N_{\mathrm{eff}, i}$ under a proposed population $\Lambda$ is
    \begin{equation}
        N_{\mathrm{eff},i}(\Lambda) \equiv \frac{\left[ \sum_{j=1}^{N_i} w_{i,j}(\Lambda)\right]^2}{\sum_{j=1}^{N_i} \left[ w_{i,j}(\Lambda)\right]^2}\,,
        \label{eq:Neff}
    \end{equation}
where $w_{i,j}(\Lambda) = p(\lambda_{i,j}|\Lambda)/p_\mathrm{pe}(\lambda_{i,j})$.
Small $N_{\mathrm{eff},i}(\Lambda)\lesssim 10$ indicates that the given event is sparsely sampled, and hence the likelihood may be dominated by sampling variance \cite{2022arXiv220400461E}. 
In this case, we should not necessarily trust the results of Monte-Carlo-based hierarchical inference.
We do not impose any cut on $N_{\mathrm{eff},i}(\Lambda)$.
Instead, we compute and track $\mathcal{N}\equiv\mathrm{min}\log_{\rm 10} \left[N_{\mathrm{eff},i}(\Lambda)\right]$ for each  model we consider.
{  In all cases we find that effective sample counts  are  large, with $\min \left[ N_{\mathrm{eff}}\right]> 10$ --
for $\pi_{\mathcal{N}+\mathcal{U}}$, $\pi_{\mathcal{N}+\mathcal{U}_w}$, 
$\pi_{\mathcal{N}+\mathcal{NU}_w}$, and
$\pi_{\mathcal{N}_m+\mathcal{U}_w}$
we find $\mathcal{N}=1.9^{+0.5}_{-0.6}$,
$1.8^{+0.5}_{-0.5}$,  $1.8^{+0.5}_{-0.4}$, 
and $1.4^{+0.6}_{-0.4}$ respectively; computing
$\mathrm{min}\left[N_{\mathrm{eff},i}(\Lambda)\right]$ for the high mass population only (i.e., 
setting $\eta(m_1)=1$ in
equation~\ref{eq:Neff}), we find 
$\mathcal{N}=2.4^{+0.3}_{-0.4}$,
$2.3^{+0.3}_{-0.3}$, $2.4^{+0.3}_{-0.4}$ and $2.0^{+0.3}_{-0.3}$. This gives us confidence that the Monte Carlo
averages over posterior samples  yields accurate results.
The posteriors of  $\mathcal{N}$ and $N_{\rm inj}$ for these models are shown in  Figs.~\ref{fig:post1},
 \ref{fig:post2},
 \ref{fig:post3}, and \ref{fig:post1b}}.

\begin{figure*}
    \centering  
    \includegraphics[width=1.\textwidth]{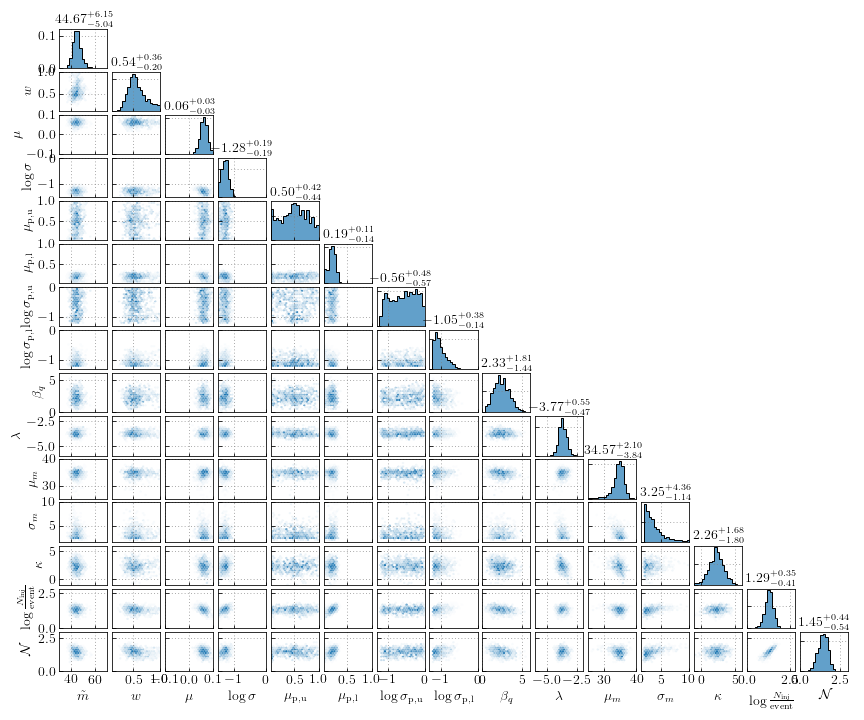}
    \caption{Posteriors on the  parameters that govern the hierarchical model where
    the $\chi_{\rm eff}$ and $\chi_{\rm p}$ distributions
    are represented using equation~\ref{eq:pi4}. Here $ \log$ indicates $\log_{10}$ and $N_{\rm inj}/event$ gives the total number of injection divided by the number of  detections.
    }
    \label{fig:postXp}
\end{figure*}

{  Particular care must be given when interpreting the results from 
the $\pi_{\chi_{\rm p}}$ model. In this model we find that setting a zero lower bound  on the
prior of $\mu_{\rm p,u(l)}$ results in an unacceptably low   $\mathcal{N}\approx 0.1$ and in a very narrow Gaussian for the $\chi_{\rm p}$ distribution of the low-mass population.
To unsure a sufficiently large sampling size and to avoid singularities, we follow 
  Ref.~\cite{2021arXiv211103634T, Abbott:2020gyp} and exclude zero from the prior range of  $\mu_{\rm p,u(l)}$ as reported in Table~\ref{tab:spin-priors}, but we caution  
 that the inference of $\chi_{\rm p}$ may
be subject to increased Monte Carlo averaging error and the assumed lower bounds for the parameter priors. We note that all other results (in particular the $\chi_{\rm eff}$ and $\tilde m$ results on which our paper is focused) are unchanged, and similar sampling issues did not present for other parameters.
 The posteriors for a selections of parameters for
the $\pi_{\chi_{\rm p}}$ model are shown in Fig.~\ref{fig:postXp}. 
}

\end{document}